# Radiative Thermal Transistor

**Yuxuan Li[1], Yongdi Dang[1], Shen Zhang[1], Xinran Li[1], Yi Jin[1], Philippe Ben-Abdallah[2*], Jianbin Xu[3] and Yungui Ma[1*]**

[1]*State Key Lab of Modern Optical Instrumentation, Centre for Optical and Electromagnetic Research, College of Optical Science and Engineering; International Research Center (Haining) for Advanced Photonics, Zhejiang University, Hangzhou, 310058, China*

[2]*Laboratoire Charles Fabry, UMR 8501, Institut d'Optique, CNRS, Université Paris-Saclay, 2 Avenue Augustin Fresnel, 91127 Palaiseau Cedex, France*

[3]*Department of Electronic Engineering, The Chinese University of Hong Kong, Shatin, Hong Kong, China*

Corresponding authors' E-mail: yungui@zju.edu.cn;
pba@institutoptique.fr

Abstract:

**Developing thermal analogues of field-effect transistor could open the door to a low-power and even zero-power communication technology working with heat rather than electricity. These solid-sate devices could also find many applications in the field of active thermal management in numerous technologies (microelectronic, building science, energy harvesting/conversion…). Recent theoretical works has suggested that a photonic transistor made with three terminals can in principle be used to switch, modulate, and even amplify heat flux through exchange of thermal photons. Here, we report an experimental demonstration of thermal transistor effect using a non-contact system composed by a temperature-controlled metal-insulator-based material interacting in far-field regime with two blackbodies held at two different temperatures. We demonstrate that, with a tiny change in the temperature of the active layer, the heat flux received by the cold blackbody can be drastically modified. An amplification parameter of heat flux over 20 is reported.**

The transistor introduced by Bardeen and Brattain [1] is a key element of modern computer technology which has revolutionized our daily life. Due to its unique capabilities to control and amplify an electric current, this element is the cornerstone of any electric logic circuit [2]. Thermal counterparts of such device which make possible the control of heat flow in a similar way as electric currents in transistor are not as widespread today. However, in 2006, Li et al. [3] have proposed a thermal analog of field effect transistor (FET) to achieve such a control exploiting acoustic phonons transport in interconnected nonlinear solid elements under the action of a temperature gradient, opening so the way to information treatment with heat [4,5]. However, these solid phononic devices suffer from some weaknesses of fundamental nature which intrinsically affect their performances. Among them, the speed of sound in solids naturally reduces the operating speed of these systems. To overcome this issue, a radiative transistor has been proposed in 2014 by Ben-Abdallah and Biehs [6] exploiting heat carried by thermal photons rather than by phonons. The radiative thermal transistor (RTT) consists of three parts, a solid source, a drain, and a gate, in analogy to a conventional FET. The source and drain are made of materials which are held at different temperatures to create a temperature gradient. The source, which is hotter than the drain, emits thermal photons which transfer heat to the drain through the gate. This photonic transistor employs a material for the gate which undergoes a phase transition around a critical temperature. By operating around this temperature, the gate can switch from a dielectric (insulator) behavior at low temperature to a metallic (conductor) behavior at high temperature. By slightly raising the temperature of the gate beyond the critical temperature, a negative differential resistance [3] (NDTR) appears in the system giving rise to an amplification of heat flux received by the drain, the so-called thermal transistor effect.

Thereafter, different versions of radiative transistors with improved controllability or performances have been proposed and theoretically studied [7-13], mostly incorporating phase change materials (PCMs). In parallel, the strong temperature dependence of PCMs has been used to rectify heat flux in two terminal systems and to generate bistable thermal states in multi-terminal systems, allowing so the development of thermal diodes [13-20] and thermal memories [21-23] as well as thermal logic gates [24,25]. However, to date, no experiment has ever been carried out to demonstrate the reality of radiative thermal transistor effect, the lack of experimental results being probably related to the difficulty to characterize heat exchanges in a many-body thermophotonic system [26,27]. Here we report an experimental study of a photonic thermal transistor working in far-field regime.

The investigated system is sketch in Figure 1 beside a classical electric FET. Our radiative thermal transistor consists of two blackbody paint (emissivity ~ 0.98) coated thermostats for source and drain and two $VO_2$ thin films coated silica substrates sandwiching a thin metallic heater for gate. The source and the drain devices are supported by two thermostats fixed at the bottom of the vacuum chamber, as shown in Fig.1 (c). After moving the gate suspended by a cantilever to a position where the three terminals are separated at equal vacuum gaps of ~0.5mm, it is also fixed in place using a screw. There are embedded heat flux sensors in the source and the drain, which are attached to the corresponding copper blocks with silicone gel. The measuring area of the sensors is identical to the radiative exchange surfaces of the system. The source and the drain are mainly made of blackbody paint coated polished copper blocks besides the inserted heat flux sensors, thermistors, thermoelectric device, and a Peltier element. The $VO_2$ is a metal-insulator transition (MIT) material [28-30] which undergo a first-order phase transition around the critical temperature $T_c$ = 340 K. This gate is intercalated between a source (hot body) and a drain (cold body) which are two thermostats made with bulk solids coated with a black paint with an emissivity close to unity in the infrared range. Therefore, these thermostats behave like two blackbodies. The temperature $T_S$ is maintained by a thermostat made of a thermoelectric device and a Peltier element, while $T_D$ is kept at a constant temperature identical to the ambient by adhering to the vacuum chamber wall. By applying a 'bias' temperature $T_G$ on the gate, the flow of heat $Q_S$ lost by the source or the flow $Q_D$ received by the drain can be dynamically controlled. The whole system itself is in interaction with the surrounding environment at temperature $T_{env} = 296 \pm 2$ K. Each part of the system exchanges heat by radiation with this external bath and by conduction through the electric wires. The experiments being carried out in a high vacuum ($\sim 10^{-4}$ Pa), convective exchanges are negligible. Moreover, the whole system is suspended in order to suppress any direct heat exchanges with a substrate. In the experimental platform, the three terminals system has identical exchange surfaces ($10 \times 10$ mm$^2$) and the adjacent blocks are separated by a vacuum gap 0.5mm thick, a thickness which guaranties that heat exchanges take place in far-field regime and that the geometrical view factors [31] between exchange surfaces are close to unity (Supplementary material [32], section I). Moreover, heat flux exchanges by radiation with the surrounding environment are negligible in comparison with that between the blocks of RTT. When the gate thickness is larger than the penetration depth of electromagnetic waves in the infrared, the gate can be considered as an opaque medium so that the source and the drain interact only with the gate. In these conditions, the heat flux exchanged by radiation between these blocks read,

$$Q_S = \sigma \varepsilon(T_G)(T_S^4 - T_G^4), \qquad (1)$$
$$Q_D = \sigma \varepsilon(T_G)(T_G^4 - T_D^4), \qquad (2)$$

where $\sigma$ denotes the Stefan-Boltzmann constant and $\varepsilon$ is the spectral average emissivity of the gate which is temperature dependent. It follows that the net radiative flux received the gate writes,

$$Q_G = Q_D - Q_S. \qquad (3)$$

The sign of $Q_G$ can be either positive (heating) or negative (cooling) depending on the temperature of terminals. In steady state regime and without external heat source applied on the gate

$$\frac{dT_G}{dt} = Q_G = 0, \qquad (4)$$

and the gate stabilizes at the equilibrium temperature

$$T_e = [0.5 \times (T_S^4 + T_D^4)]^{1/4}. \qquad (5)$$

To get an operating transistor, $T_e$ must be set in the vicinity of the critical temperature $T_c$ of MIT so that a small perturbation of the gate temperature can trigger the phase transition of this material and consequently significantly modify the heat flux flowing through the transistor. The source and drain being blackbodies, the variation of this flux in the transistor is simply driven by the emissivity difference $\Delta\varepsilon$ of the gate between the situation where the MIT material is insulating and the situation where it is metallic. This amplitude has been maximized by optimizing the thickness $d$ of $VO_2$ film deposited on the $SiO_2$ substrate. For assessing this variation, the dielectric properties of $VO_2$ have been taken from the literature [28,33] (Supplementary material [32], section II). In the transition region between the purely dielectric and the purely metallic phase, the dielectric properties of $VO_2$ film are described from the Looyenga mixing rule [34-36]. As for the optical constant of silica substrate, we used the database coming from the literature [37,38]. Regarding the spectral average emissivity of the gate, it is related to spectral emissivity $\varepsilon_\lambda$ as follows:

$$\varepsilon(T_G) = \frac{\int_0^{+\infty} I_\lambda^0(T_G)\varepsilon_\lambda(T_G)d\lambda}{\sigma T_G^4}, \qquad (6)$$

where $I_\lambda^0(T)$ is the blackbody radiation intensity at temperature $T$ and wavelength $\lambda$. An optimal thickness of 300 nm for the $VO_2$ film has been chosen in order to maximize the radiative switch induced by the phase transition of this material (Supplementary material [32], section III). Silica substrates with a thickness of 500 µm and an area of $10 \times 10$ mm$^2$ were cleaned by ultrasonic waves sequentially in acetone, methyl alcohol, and isopropyl alcohol. Each step was for 5 min. About 300-nm thick vanadium dioxide ($VO_2$) film was deposited on the clean silica substrates with vanadium target by magnetron sputtering with DC power of 200W. During deposition, the chamber pressure was maintained at 5.5 mTorr with an $Ar/O_2$ gas mixture at a flow rate of 70/4 sccm. The samples are later heated to 450 °C for the formation of the $VO_2$ phase.

With this geometric configuration an emissivity contrast between the two phases of $\Delta\varepsilon \sim 0.5$ can be achieved. Figure S4(a) plots the temperature-dependent emissivity of the gate sample measured in both heating and cooling processes, derived from the reflectivity spectra (inset) obtained via a Fourier Transform Infrared Spectrometer (FTIR). The measurement of the gate emissivity with respect to its temperature have demonstrated the hysteric behavior of the gate around the critical temperature of $VO_2$, the phase transition of $VO_2$ film starting at $T_{c1} = 340.4$ K and being fully achieved at $T_{c2} = 350$ K with a thermal hysteretic width $\Delta T \approx 10.8$ K. A maximum derivative slope ($\sim$ -0.065 K$^{-1}$) for the $\Delta\varepsilon(T)$ curve appears at 346.5 K, which is vital in enhancing the amplification effect of the transistor. Figure S4(b) plots the measured and calculated bi-state emissivity spectra of the gate sample. They show good agreement. The slight difference might be caused by the nonstoichiometry and granularity of the $VO_2$ film.

To describe the operating regimes of transistor we show in Figure 2(a) the measured (symbols) heat flux $Q_S$, $Q_G$ and $Q_D$ of transistor with respect to the control temperature $T_G$ of the gate when $T_S = 357.7$ K and $T_D = 298.1$ K. The temperature of the source, the gate and the drain are all measured by thermistors inserted into them. Heat flux lost/received by the source and the drain are measured using embedded sensors (HS-10, Captec Enterprise). The conversion from voltage recorded by a digital source meter (2450, Keithley) to heat flux is determined by the calibration curve provided by the HFS manufacturer. The raw data of $Q_S$ and $Q_D$ are all recorded after reaching thermal equilibrium. The calibration of background heat loss and the thermal resistance network corresponding to the experimental configurations are described in details in section V of Supplementary material [32]. These measures are compared with the theoretical predictions, plotted in dashed lines, and clearly demonstrate a good agreement with them. For the investigated configuration, the equilibrium temperature which corresponds to the temperature of the gate at the intersection of $Q_S(T_G)$ and $Q_D(T_G)$ curves, is $T_e = 331.9\ K$, a temperature which is smaller than the critical temperature $T_c$=340 K of MIT material. By increasing $T_G$ by less than 10 degrees the transistor operates in the transition of MIT material. In this region, the net flux curves $Q_G(T_G)$ shows two extrema values at $T_{p1} = 345$ K and $T_{p2} = 349$ K. Between these two temperatures, the slope of $Q_S(T_G)$ and $Q_D(T_G)$ curves are both negative demonstrating so the presence of a radiative NDTR [39] (see inset of Fig. 2(a)). Analogously to its electronic counterpart, this feature is at the origin of transistor effect.

In a similar way to FET, the RTT can act as a switch, a modulator, or an amplifier. These functionalities are described below.

***Thermal switching.*** The transistor can act like a good heat conductor or an insulator depending on the gate temperature $T_G$. Following the conventional definition [39], when the flux received by the drain is minimal, the transistor is said to operate in a "off" mode. On the opposite, when this flux reaches its maximal value, the transistor operates in "on" mode. This, switch between the "on" and "off" modes takes place (see Fig. 2(a)), in the transition region, when the gate temperature goes from $T_G = T_{G,max}$ to $T_G = T_{G,min}$, $T_{G,max}$ and $T_{G,min}$ being the gate temperatures leading to a minimal and maximal value of $Q_D$. Between these two temperatures, the transistor is "semi-on". For our device, the heat flux $Q_S$ and $Q_D$ are reduced by 110.8 and 92.2 W/m$^2$ respectively through this switching operation with a change of about ten degrees of the gate temperature. If this temperature varies is beyond or below the transition region, then according to the non-monoticity of flux curves with respect to $T_G$, the gate follows a hysteric trajectory during its heating and cooling process. By analogy with the FET, the switching ratio for the transistor is defined by the parameter

$$\eta_{S,D} = \frac{Q_{S,D}(T_{G,max})}{Q_{S,D}(T_{G,min})}. \qquad (7)$$

The calculations indicate that $\eta_i$ is inversely proportional to the variation of the gate temperature (i.e. $\eta_i \propto \gamma \cdot \Delta T_G^{-1}$, $\Delta T_G = T_{c2} - T_{c1}$ being the temperature interval where the phase-transition takes place and $\gamma = \varepsilon^d/\varepsilon^m$ being the dielectric-to-metallic ratio). The results plotted in Fig. 2(a), lead to switching ratio $\eta_S = 14.3$ and $\eta_D = 1.7$, corresponding to a power variation received by the drain of the order of 20 W/m$^2$ when $T_S = 357.7$ K and $T_D = 298.1$ K. Notice that this on/off ratio is significantly larger than the previously reported values in phononic thermal switches [40,41], and is comparable to an electrochemical thermal transistor [42]. Recently, a three-terminal magnetic thermal transistor with source-drain thermal switch ratios of near 100 has been reported, as it directly provides or breaks thermal contact through a magnetic force-controlled shuttle [43]. Moreover, with our radiative device, the switching time can in principle be relatively small. Although, the phase transition of VO$_2$ can take place at sub-picosecond scale [29,44,45], the thermal inertia intrinsically limits the operating speed of transistor. With macroscopic system, this timescale is typically of the order of the second up to few minutes. But this time can reduce to few milliseconds with nanostructures and could even be much smaller with two-dimensional materials [46].

***Thermal modulation.*** Thermal modulation can be achieved with the RTT by oscillating the gate temperature around the phase transition region. Note that while the variation of the heat flux $Q_G$ remains relatively small under the action of this oscillation (Fig. 2(a)), the heat flux $Q_S$ and $Q_D$ can be intensely modulated by setting $T_e < T_{c1}$ (a much larger modulation can even be realized in near-field regime [6]). The measured results indicate that our transistor can modulate the heat flux $Q_S$ and $Q_D$ with a relatively large amplitude, over 100 W/m$^2$ with a 10K temperature change of the gate. This variation corresponds to the change of flux radiated by a blackbody when its temperature goes from $T_c$ to $T_c + \Delta T$ with $\Delta T = 10\ K$.

***Thermal amplification.*** The heat flux amplification associated to the transistor can be quantitatively evaluated by the two parameter [3,6]:

$$\alpha_{S,D} = \left| \frac{\partial Q_{S,D}}{\partial Q_G} \right| = \left| \frac{1}{R_{S,D}} \frac{R_S R_D}{R_S + R_D} \right|, \quad (8)$$

where $R_S = -\left( \frac{\partial Q_S}{\partial T_G} \right)^{-1}$ and $R_D = \left( \frac{\partial Q_D}{\partial T_G} \right)^{-1}$ denote the differential thermal resistances between the source and the gate and between the gate and the drain. It turns out, according to expression (8), that the existence of a transistor effect (i.e., $\alpha_{S,D} > 1$) requires the presence of a NDTR. It is worthwhile to note that the rhs of expression (8) shows that this condition is necessary but it is not sufficient to guaranty the presence of transistor effect. The variation of the RTT amplification factor is plotted in Fig. 2(b) under different operating temperatures. When $T_G$ is set outside the phase transition region we see in Fig. 2(a) that the $Q_S(T_G)$ and $Q_D(T_G)$ curves have slopes with an opposite sign and the same absolute value. It turns out that, $R_S = R_D$ and $\alpha_{S,D} = 0.5$. On the other hand, inside the transition region, these curves have negative slopes so that $R_S > 0$ and $R_D < 0$. In Fig. 2(b) we see that $\alpha_{S,D} > 1$. In particular, when $T_S = 357.7$ K and $T_D = 298.1$ K two prominent peaks of amplification appear with a magnitude $\alpha_D = 17.5$ and $\alpha_D = 22.6$ respectively at $T_G = 346$ K and $T_G = 349$ K, two temperatures which are very close to the temperatures where $Q_G$ passes through its extremal values as shown in the inset of Fig. 2(a). This behavior can be analytically described from expression (8) using the definitions (1-3) of flux $Q_i$ written with respect to the gate and equilibrium temperatures. It is then straightforward to show that the operating temperature $T_G$ which maximize the amplification factor satisfies the following equation [9]:

$$\varepsilon'(T_G)(T_G^4 - T_e^4) + 4T_G^3 \varepsilon(T_G) = 0, \quad (9)$$

where $\varepsilon'(T) = d\varepsilon/dT$. Therefore, given an equilibrium temperature $T_e$ (set by the two thermostats temperatures $T_S$ and $T_D$). the 'bias' temperature which maximizes the amplification factor can be identified by solving this equation. However, due to the bistable behavior of the system, as shown in Fig. 2(b), this solution is not unique. In our experiment, for $T_e$ = 331.9 K these solutions are $T_G$ = 343.9 K and 350.3 K.

A complete description of physical characteristics of RTT with respect to the input temperatures has been carried out. The results are plotted in Figs. 3(a)-3(d). In Fig.3(a), the amplification coefficient $\alpha_D$ of transistor (i.e. amplification of flux received by the drain) is mapped with respect to $T_e$ and $T_G$ when the cold thermostat is set at $T_D$ = 300 K. We observe that this coefficient can be about four orders of magnitude ($\alpha_D \sim 2.2 \times 10^4$) larger than the net flux injected in the gate in specific operating zones (yellow bright area of Fig. 3(a)), revealing the potential of such RTT to operate with weak flux. Moreover, we see that this coefficient can be drastically modified by tiny changes on the gate temperature. This demonstrates that by working in narrow temperature range we can easily modulate or amplify the flux received by the drain. Moreover, for such transistor we see from Figs. 2(a), 3(b) and 3(c), that the transistor effect becomes prominent mainly when $T_e \leq T_c$. Beyond the critical temperature of MIT material, the flux amplification tends to disappear as shown in Fig. 3(d). This evolution is directly related to the change of differential thermal resistances with respect to the gate temperature. As show the comparison of flux curves plotted in Figs. 3(b) and 3(c), the slope of the $Q_S(T_G)$ curve strongly increases with the source temperature $T_S$ while the slope of $Q_D(T_G)$ is relatively insensitive to it in the region where the phase transition takes place. It turns out that $R_S$ increases with $T_S$ while $R_D$ remains approximately constant. This evolution of differential resistances induces, according to (8), a decrease in the amplification coefficient. Fig. 3(a) also clearly shows the existence of two possible optimal values for the gate temperature that maximizes the transistor effect, given $T_e$ or $T_S$. This property which is due to the temperature dependence of the gate emissivity is rarely observed in electronic FETs.

Inspection of expression (8) shows that the transistor effect can also be observed in a system which displays both a negative resistance $R_S$ and a positive resistance $R_D$. This can actually come true with a gate made with a VO$_2$-based metasurface. This metasurface is schematically drawn in Fig. 4(a). It is composed on the top by VO$_2$ circular pellets (radius $r$ = 0.85 µm, thickness $h_2$ = 0.2 µm) arranged in square periodic lattice of period $p$ = 3 µm. This lattice is deposited on a highly-doped silicon layer of thickness $h_1$ = 0.8 µm which itself covers a gold layer ($h_0$ = 0.5 µm) which can be used to control the gate temperature. The optical response of the metasurface is calculated using a finite-element method [47] and its emissivity (blue lines) in Fig. 4(b) is calculated using the Kirchhoff's law of thermal radiation. As previously, in the phase-transition region, the Looyenga mixing rule is used to calculate the emissivity. Unlike the gates made with VO$_2$ films (red lines) in Fig. 4(b), such a structure gives rise to low mean emissivity at low temperature when the MIT material is in its dielectric phase and to an intense emissivity at high temperature when VO$_2$ becomes amorphous. It turns out an inversion of the sign of differential thermal resistances of RTT made with this gate in comparison with a transistor made with a film of VO$_2$. This change is clearly visible in Fig. 4(c) when $T_S = 380$ K and $T_D = 300$ K. One of the advantages of this type of transistor is that it can operate at lower temperature, making so wider the range of temperatures where the transistor effect can be used using a given MIT material. The full characteristics curve of this transistor is plotted in Fig. 4(d). In a similar way as for the transistors based on VO2-flms, this map shows that this transistor can achieve a strong heat flux amplification. But unlike the formers, we see that this transistor works by cooling down the gate, the equilibrium temperature being greater than $T_{c2}$ (see Fig. 4(c)).

In conclusion, we have experimentally analyzed a far-field radiative thermal transistor and highlighted its basic operating modes demonstrating so the versatility of these multi-terminal elements to switch, modulate or amplify heat flux. The thermal transistor effect has been unambiguously demonstrated and an amplification factor of flux greater than 20 has been measured. We think that this building block paves the way to a new generation of devices for active thermal management, to innovative wireless sensors using heat as a primary source of energy and to a 'low-electricity' technology enabling information processing. In these devices, the infrared emission coming from various systems (people, machines, electric devices…) which results from dissipating processes could be captured by thermal transistors to launch a sequence of logical operations in order to control the heat propagation, to trigger specific actions (thermo-mechanical coupling with MEMS, thermal energy storage…) or even process information (Boolean treatment of heat). This technology, could provide a new degree of freedom in the direct interaction of machines and smart systems with their environment using thermal signals, ideally without external electricity supplying. By embedding pervasive intelligence and sensing based on purely thermal signals into machines, a higher level of awareness and automation could be implemented in a near future. Hence, the development of thermal circuits could open the door to a lower-power technology for the Internet of Things by allowing machine-to-machine communication with heat.

**Acknowledgements**

YGM thanks the partial supports from National Natural Science Foundation of China 62075196, Natural Science Foundation of Zhejiang Province LXZ22F050001 and DT23F050006, and Fundamental Research Funds for the Central University (2021FZZX001-07). JBX would like to thank RGC for support via AoE/P-701/20.

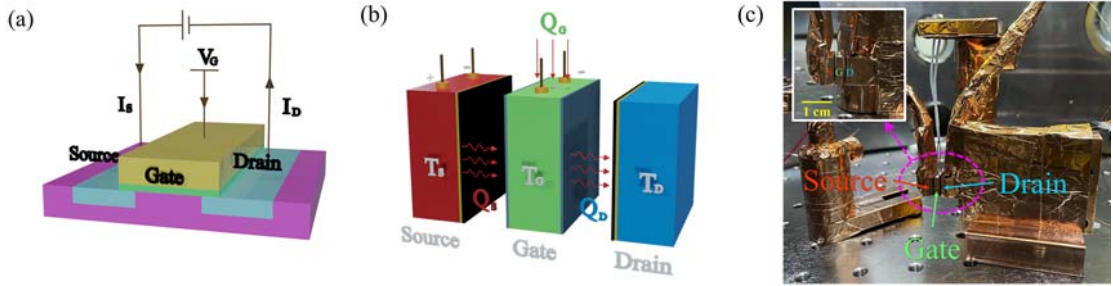

FIG. 1. Sketch of field effect transistor and radiative thermal transistor. (a) A FET is a three-terminal system composed by a source, a gate and a drain which correspond to the emitter, base, and collector of electrons, respectively. By applying an external bias voltage $V_G$ on the gate, the electric current $I_D$ flowing between the source and the drain can be switched, modulated or amplified. (b) A RTT is composed with a solid layer made with a phase-change material and which is placed between two thermal emitters held at different temperatures $T_S$ and $T_D$ at a separation distance which is much larger than their thermal wavelength. By changing the temperature $T_G$ of the gate around the critical temperature of PCM using external heaters (electrodes), the heat flux $Q_S$ (resp. $Q_D$) exchanged by radiation between the source and the gate (resp. between the gate and the drain) can be switched, modulated, and even amplified. (c) Experimental setup of the RTT.

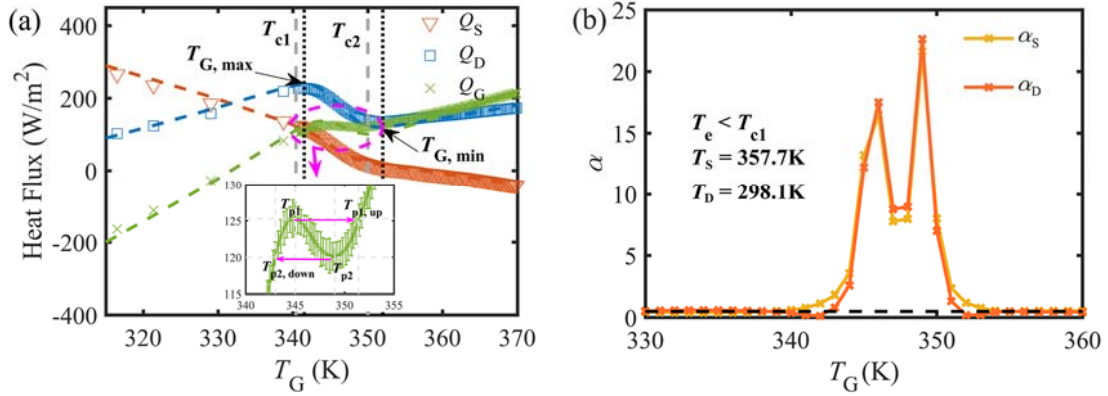

FIG. 2. Characteristic curves of thermal transistor. (a) Measured (symbols) and calculated (dashed lines) internal radiative heat flux $Q_S$ (resp. $Q_D$) between the source and the gate (resp. between the gate and the drain) and the net flux $Q_G$ on the gate from the gate when $T_e < T_{c1}$. The dashed vertical lines, identified by emissivity measurement, represent the region where the phase transition occurs. This region is enlarged in the inset. The error bars representing the standard deviation had been added in the measured data, which can be clearly seen from the zoom-in curves in the inset. (b) Amplification factors $\alpha_{S, D}$ of transistor when the temperature of the source and drain are held at $T_S = 357.7$K and $T_D = 298.1$K, respectively. In the transition region, $\alpha_{S, D} > 1$ owing to the presence of a negative differential thermal resistance $R_D$. The existence of two solutions is related to the bistability of transistor.

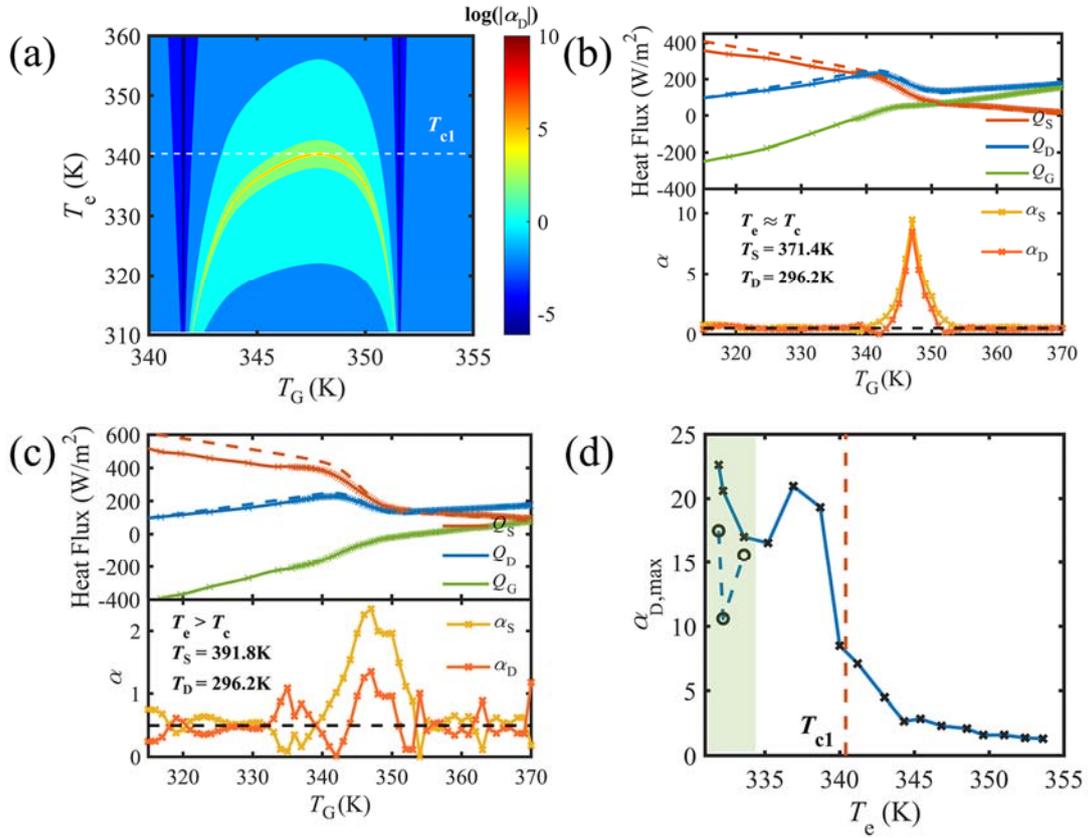

FIG. 3. Influence of input temperatures on physical characteristics of RTT. (a) Mapping of amplification factor $\alpha_D$ with respect to $T_e$ and $T_G$ when $T_D = 300$ K. Note $\alpha_D$ takes a logarithmic scale. (b, c) Measured (symbols) and calculated (dashed lines) heat flux (top) and the corresponding amplification factors $\alpha_{S, D}$ (bottom) as a function of $T_G$ when $T_e \approx T_{c1}$ and $T_e > T_{c1}$, respectively. (d) Maximum measured value $\alpha_{D,max}$ of amplification factor for different equilibrium temperature $T_e$. The open circles and the x's indicate the two extreme values of amplification factor when $T_e$ is much lower than $T_{c1}$.

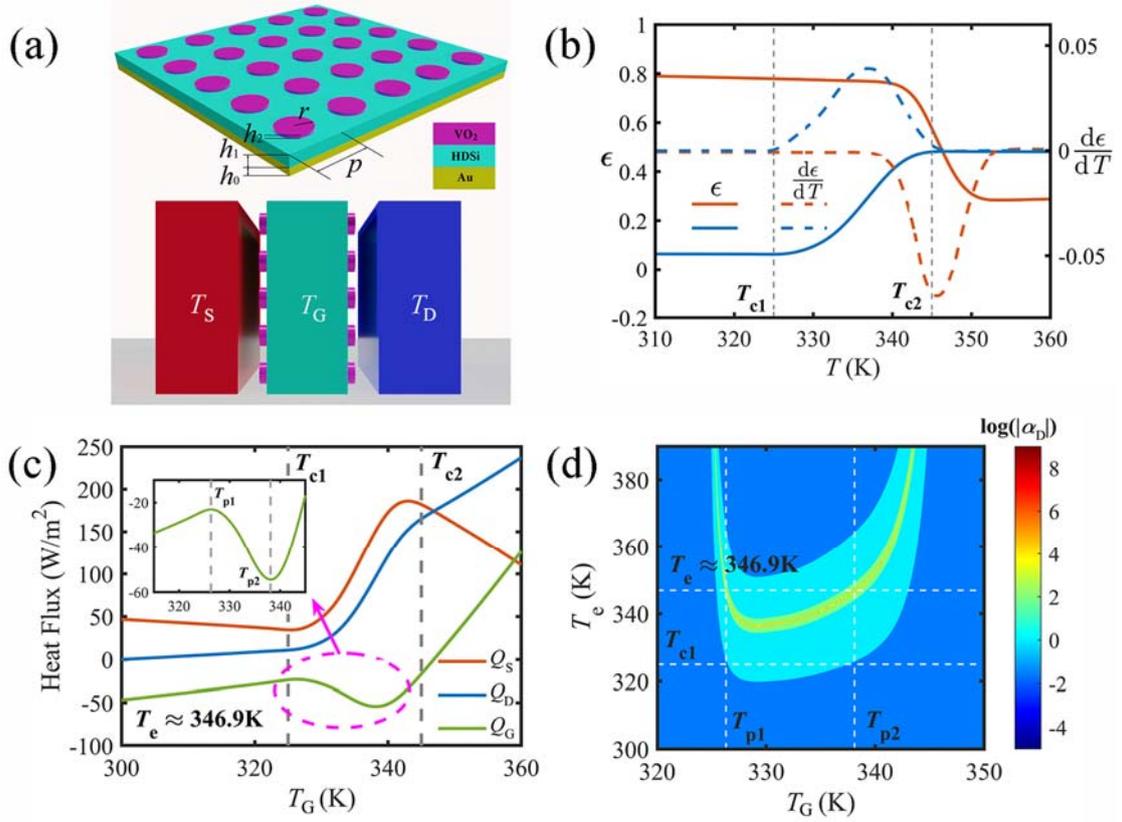

FIG. 4. Thermal transistor based on a gate with positive differential thermal emissivity. (a) Schematic representation of the metasurface (upper) used to design the gate and transistor (lower). (b) Spectral average emissivity of the metasurface (blue) and of a VO$_2$ film (red) with respect to the temperature. Dashed curves show the derivative of emisivities with respect to the temperature. (c) Heat flux exchanged in the transistor with respect to the gate temperature when $T_e \approx 346.9$ K. The inset details the region where $R_D < 0$. ($T_{p1} = 326.3$ and $T_{p2} = 338.1$ K). (d) Mapping of amplification factor $\alpha_D$ with respect to $T_e$ and $T_G$ when $T_D = 300$ K.